\documentclass[aps,prl,twocolumn,showpacs,superscriptaddress]{revtex4}

\usepackage{graphicx}         
\usepackage{bm}               
\usepackage{amssymb}          
\usepackage{amsmath}          

\def \msun        {M_\odot}

\begin{document}
\title{Recognising Axionic Dark Matter by Compton and de-Broglie Scale Modulation of Pulsar Timing.} 

\author{Ivan De Martino }
\email{ivan.demartino@ehu.eus}
\affiliation{Department of Theoretical Physics, University of the Basque Country UPV/EHU,
             E-48080 Bilbao, Spain}

\author{Tom Broadhurst}
\email{tom.j.broadhurst@gmail.com}
\affiliation{Department of Theoretical Physics, University of the Basque Country UPV/EHU,
             E-48080 Bilbao, Spain}
\affiliation{Ikerbasque, Basque Foundation for Science, E-48011 Bilbao, Spain}

\author{S.-H. Henry Tye }
\email{iastye@ust.hk}
\affiliation{Institute for Advanced Study and Department of Physics, Hong Kong University of Science and Technology, Hong Kong}

\author{Tzihong Chiueh}
\email{chiuehth@phys.ntu.edu.tw}
\affiliation{Department of Physics, National Taiwan University, Taipei 10617, Taiwan}
\affiliation{National Center for Theoretical Sciences, National Taiwan University, Taipei 10617, Taiwan}

\author{Hsi-Yu Schive}
\email{hyschive@gmail.com}
\affiliation{National Center for Supercomputing Applications, Urbana, IL 61801, USA}
              
\author{Ruth Lazkoz}
\email{ruth.lazkoz@ehu.eus}
\affiliation{Department of Theoretical Physics, University of the Basque Country UPV/EHU,
             E-48080 Bilbao, Spain}

\date{\today}


\begin{abstract}

Light Axionic Dark Matter, motivated by string theory, is increasingly favored for the   ``no-WIMP era". 
Galaxy formation is suppressed below a Jeans scale, of $\simeq 10^8 M_\odot$ by 
setting the axion mass to, $m_B \sim 10^{-22}$eV, and the large dark cores of dwarf galaxies are 
explained as solitons on the de-Broglie scale. This is persuasive, but detection of the inherent scalar 
field oscillation at the Compton frequency, $\omega_B= (2.5{\rm \, months})^{-1}(m_B/10^{-22}eV)$, 
would be definitive. By evolving the coupled Schr\"{o}dinger-Poisson equation for a  Bose-Einstein 
condensate, we predict the dark matter is fully modulated by de-Broglie interference, with a dense 
soliton core of size $\simeq 150pc$, at the Galactic center. The oscillating field pressure 
induces General Relativistic time dilation in proportion to the local dark matter density and pulsars within 
this dense core have detectably large timing residuals, of $\simeq 400nsec/(m_B/10^{-22}eV)$. This is encouraging as 
many new pulsars should be discovered near the Galactic center with planned radio surveys.
More generally, over the whole Galaxy, differences in dark matter density between pairs of pulsars imprints a 
pairwise Galactocentric signature that can be distinguished from an isotropic gravitational  wave background. 

\end{abstract}

\maketitle

\section{Introduction}

Axions are a compelling choice for extending the standard model in particle physics, generating oscillating 
dark matter with symmetry  broken by the misalignment mechanism \cite{Preskill:1982cy,Abbott:1982af,Dine:1982ah}. 
Such fields are generic to string  theory, arising from dynamical compactification to 4 space-time 
dimensions,  so that multiple stabilized, complex scalar fields naturally appear (e. g. Ref \cite{Svrcek:2006yi}).
Such an axion is effectively massless until the Universe cools below some critical temperature, and rolls down a small non-pertubatively generated potential, oscillating about the minimum, corresponding to a coherent zero-momentum axion.
It is argued that very light axions are very natural in this context \cite{Hui:2016ltb} and furthermore, 
in a ''string theory landscape", the cosmological constant may also be naturally small and accompanied 
by correspondingly very light scalar bosons \cite{Tye:2016jzi}. 

Axion oscillation generates field pressure at frequency $2m_B$, with static energy density to leading order that is  coherent below the de-Broglie scale. At a certain scale, self-gravity is balanced by field pressure, yielding a static, centrally located density peak, or soliton. The soliton scale depends on the gravitational potential corresponding to
only 150 pc for the Milky Way \cite{Schive2014}. Throughout the Galaxy, gravity is balanced by the axion 
pressure forming density ``granules" 
of similar scale to the soliton from interference of many 
large-amplitude de Broglie waves which are unstable, regenerating themselves with a lifetime of 
about 1 Myr for our Galaxy \cite{Schive2014b}.

Khmeinitsky and Rubakov (2013) made the very interesting observation that the field pressure oscillations on the Compton 
scale leads to an oscillating gravitational potential that can affect pulsar timing measurements \cite{Khmeinitsky:2013lxt}. 
They assume a constant local dark matter density so pulsar timing and our Earth clock will be affected in the same way, with a net relative timing modulation from the phase difference that will be very challenging to detect. However, we show here that a much stronger signal is expected towards the Galactic center, in and around the solitonic core that we predict for 
light axionic Dark Matter. This additional de-Broglie structure is absent in \cite{Khmeinitsky:2013lxt}, which predates 
the first simulations of galaxies in this context, \cite{Schive2014}. We show this greatly enhances the expected pulsar 
timing modulation by 2-3 orders of magnitude within the central Kiloparsec of our Galaxy, for our favoured axion mass 
($0.8\times 10^{-22}eV$, \cite{Schive2014}), well 
within the current bounds of detectability, providing a direct, practical test for Axions.

Here we calculate this additional de-Broglie effect self consistently, by including our simulated halo structure. 
This is an important feature that is absent in Ref \cite{Khmeinitsky:2013lxt}, which if observed would directly 
support light bosonic dark matter.  The parameters assumed are: Hubble constant $H_0=70$ km/sec/Mpc with critical 
density $\rho_c = 6\rm{\, GeV \, m}^{-3}$, and present dark matter density $\rho_0 = 1.5 \rm{\, GeV \, m}^{-3}$. 

\section{Compton Scale Pressure Oscillation and Pulsar Timing}

For a single real scalar field $\phi$ with harmonic potential oscillation:
$\phi({\bf x},t) = A({\bf x}) \cos( m_B\,t + \alpha ({\bf x}))$, where $m_B$ is the associated boson mass.  We let $\Re[\psi]\cos(m_B t)+\Im[\psi]\sin(m_B t)\equiv m_B^{1/2}\phi$ to relate $\phi$ to the complex wave function of  Schroedinger's equation, so $m_B^{1/2}A$ and $\alpha$ are the amplitude and phase of $\psi$, respectively. From the energy-momentum tensor one obtains an oscillating pressure $p({\bf x},t) = - \frac{1}{2}m_B^2 A^2 \cos{(\omega\,t + 2 \alpha)}$ 
with frequency $\omega =2\pi \nu =2 m_B$. These oscillations are usually neglected as
the average pressure over the period is zero. However, at the Compton scale 
of interest here, the oscillating scalar field generates an oscillating gravitational potential in proportion 
to the local mass density \cite{Khmeinitsky:2013lxt}. All clocks are modulated by such an oscillating field, 
including Earth clocks and Pulsars.

\begin{figure}[t]
\centering
\includegraphics[width=8.0cm]{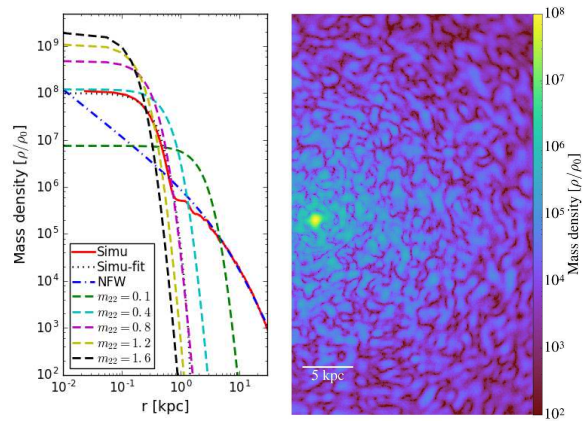}
\caption{The left panel shows predicted soliton profiles for a range of $m_B$ (dashed curves), 
including the profile of a massive simulated halo of $10^{11}\msun$ (solid red curve), corresponding to the simulation on the right (using the \texttt{yt} package \cite{Turk2011}), where
the granular de-Broglie scale structure is visible on all scales, 
including the dense central soliton. Also indicated is an NFW profile (blue dashed curve) that fits well the azimuthally averaged galaxy profile (solid red curve). 
}\label{fig1}
\end{figure}

It is customary to write the time dependent part of the time residuals as 
the relative frequency shift of the pulse 
\begin{equation}\label{eq:res}
\delta t(t) = - \int_0^t \frac{\nu(t') - \nu_0}{\nu_0} \,d t',
\end{equation}
where $\nu(t')$ is the pulse frequency at the detector, emitted at a distance $D$ at the time $t'= t - D/c$, and
$\nu_0$ is the pulsar emission frequency. To rewrite the Eq. \eqref{eq:res} in terms of the oscillating
gravitational potential we consider the linearized Einstein equations in the Newtonian gauge with two
potentials $h_{00}=2\Phi$ and $h_{ij}=-2\Psi\delta_{ij}$. Then, the frequency shift is \cite{Gorbunov2011}
\begin{equation}\label{eq:SW}
\frac{\nu(t) - \nu_0}{\nu_0} \approx \Psi(\mathbf{x_p},0)- \Psi(\mathbf{x},t).
\end{equation}
where $\bf{x_p}$ is the source position.
Thus, to compute the frequency shift we need only consider the oscillating contribution to $\Psi$. Therefore,
from the spatial components of the linearized Einstein equations the amplitude is:
\begin{equation}\label{eq:Psi_c}
 \Psi_c({\bf x}) =  \pi\frac{G\rho_{DM}({\bf x})}{m_B^2}.
\end{equation}
We stress the amplitude of the oscillation will vary over the Galaxy density with large de-Broglie scale modulations about the mean which is close to the Navarro-Frenk-White (NFW) profile \cite{NFW1996}
(see Figure \ref{fig1}). 

Plugging eqs. \eqref{eq:SW} and \eqref{eq:Psi_c} into \eqref{eq:res}, we obtain the time-dependent part of the time residuals for the 
$i-th$ pulsar with respect to the average signal ($\Delta t_i = \delta t_i-<\delta t_i>$):
\begin{eqnarray}\label{eq:dti}
\Delta t_i(t) &=& \frac{1}{\omega}\biggl[\Psi({\bf x_i})\sin\left(\omega (t - \frac{D_i}{c})+2\alpha_i\right) - \nonumber\\
& - & \Psi({\bf x_e})\sin\left(\omega t +2\alpha_e\right)\biggr],
\end{eqnarray}
where $\alpha_i\equiv\alpha({\bf x_i})$ and $D_i$ are the phase and the pulsar distance and 
$\alpha_e\equiv\alpha({\bf x_e})$ is the phase of the Earth clock, hence
cancellation of timing residuals is possible, and when pairs of clocks/pulsars are maximally out of
phase their relative timing residual is enhanced, see Figure 2.

The timing differences between two well separated pulsars will be enhanced by the
modulation of the axion density field on the de-Broglie scale throughout the Galaxy, as shown in Figures 1\& 2,
as the density structure is predicted by our simulations to be fully 
modulated on the de-Broglie scale.


\section{\MakeLowercase{de}-Broglie Scale Galactic Structure}

Bosonic Dark Matter, such as Axions, if sufficiently light can satisfy the ground state condition, 
where the de-Broglie wavelength exceeds the mean particle separation set by the density of dark matter. 
This is simply described by a coupled Schroedinger-Poisson equation, analogous to the Gross-Pitaeviski 
equations for a Bose-Einstein condensate. Expressed in comoving coordinates we have:
\begin{align}
 & \biggl[i\frac{\partial}{\partial \tau} + \frac{\nabla^2}{2} - aV\biggr]\psi=0\,,\\
 & \nabla^2 V =4\pi(|\psi|^2-1)\,,
\end{align}
 where $\psi$ is the wave function, $V$ is the gravitation potential and $a$ is the cosmological scale factor. The system is normalized
 to the time scale $d\tau=\chi^{1/2} a^{-2}dt$, and 
 to the scale length $ \xi = \chi^{1/4} (m_B/\hbar)^{1/2} {\mathbf x}$, where $\chi=\frac{3}{2}H_0^2 \Omega_0$  where $\Omega_0$ is the current density parameter \cite{Widrow1993}. 

The simplest ``Fuzzy Dark Matter" case of no self-interaction was first advocated  by Hu et al. \cite{Hu2000} 
for which the boson mass is the only free parameter, with further work 
in relation to dwarf galaxies \cite{Marsh2014, Bozek2015}. New Cosmological simulations in this context, dubbed $\psi_{DM}$, 
by \cite{Schive2014} have uncovered rich non-linear structure by solving the above equation, 
evolved from an initial standard power spectrum truncated at the inherent Jeans scale. A solitonic core forms within each virialized halo, naturally explaining dark matter dominated cores of dwarf spheroidal galaxies \cite{Schive2014}. The central soliton is surrounded by an extended halo with  `` granular" texture on the de-Broglie scale in Figure 1, which on average follows the NFW form outside the soliton \cite{Schive2014,Schive2014b}, as seen in Figure 1. This 
agreement can be understood because the pressure from the uncertainty principle is limited to the de-Broglie radius, beyond which is it negligible and behaves as collisionless CDM. 

The identification of the centrally stable soliton with the large cores of dSph galaxies has 
allowed the boson mass to be estimated  with little model dependence \cite{Schive2014, Schive2014b, Zhang2017}.  The best constraint comes form the well 
studied Fornax dwarf spheroidal (dSph), with an estimated halo mass is $4 \times 10^9 \msun$, 
yielding the soliton peak density is $2$  GeV cm$^{-3}$ and core width $1$ kpc \cite{Schive2014}
and $m_B=0.8\times10^{-22}$ eV, with somewhat larger $m_B$ derived using 
dSph galaxies from the SDSS survey \cite{Calabrese2016}. This allows us to predict the 
soliton scale expected for the Milky Way by using the scaling between Halo mass and Soliton mass scaling law:
$\rho_{peak} \propto M_{halo}^{4/3}$ and $r_c \propto M_{halo}^{-1/3}$
derived from simulations \cite{Schive2014,Schive2014b}, which for our Galaxy with a mass  $M_{halo}^{MW} = 2 \times 10^{12}\msun$ 
\cite{Portail2017}, implies a MW soliton peak density $\rho_{peak}^{MW}=8 \times 10^3$ GeV cm$^{-3}$ and soliton core width $r_c^{MW}=120$ pc (Figure 1).  
The Milky Way halo is generally taken to have a dark matter density $0.3$ GeV cm$^{-3}$ in the solar neighborhood, with recent careful studies by Portail et al. \cite{Portail2017}  and Sivertsson et al (2017) , revising upward this figure to $0.5--0.6$ GeV cm$^{-3}$ . Our predicted soliton for our Galaxy ranges over $4-8\times 10^3$  and  from Eq.(5) above, 
we have $\Delta t = \pi G\rho_{DM}/2 m_B^3 = (0.7-1.4)\times 10^{-27} m_B^{-3}$ sec$^{-2}$. 
Since $m_B=1.5 \times 10^{-7}$ sec$^{-1}$,  we arrive at $\Delta t = 200-400$ nsec; 
this $\Delta t$ is within the reach of current pulsar time arrays. 
 The central soliton then provides approximately two orders of magnitude enhancement of the DM density within 
 $r\lesssim 100pc$, dominating the Earth clock and pulsars elsewhere in the galaxy as shown in Figure 2 (top left).  For pairs of pulsars within this  soliton region, separated by more than the Compton wavelength of ($>pc$ scale),  the combined amplitude cancels or adds constructively to to a factor of two, as shown in Figure 2. The above assumes $m_B=10^{-22}$ eV, but allowing this to vary we have $\rho \propto m_B^2$ and $r_c \propto m_B^{-1}$, 
and thus $\Delta t \propto m_B^{-1}$, increasing the soliton density becomes higher, but the core radius is reduced.

\section{Pairwise Timing Amplitudes}

It is important to appreciate that all clocks are modulated within an oscillating scalar field {\it including Earth clocks} and so
in practice we can work only with relative timing residuals, either between pairs of pulsars or between any pulsar and a time 
standard based on precise Earth clocks. The ticks of an individual clock are cyclically slowed and increased at the Compton 
frequency, with a magnitude that is proportional to the mass density of the scalar field local to each pulsar, where the 
''amplitude" of this effect is the time difference induced by Eq. 3. In practice, pulsar timing measurements are typically 
averaged over a sizeable number of pulses on a relatively short timescale of hours and this rate is then compared 
on longer timescales, a practice that is well suited to the  larger than monthly Compton frequency modulation that we seek, set by the axion mass. This timescale, note, is unaffected by the slow changes in structure on the de-Broglie scale of $\simeq 1 Myr$
that set the amplitude of the timing residual for a given axion mass via equation 3.

The timing amplitude of any such modulation is determined by the de-Broglie scale density modulation which is largest for a pulsars within the central  
soliton and also when the pulsars are spatially located such that they are out of phase relative to the Compton frequency. 
In general this phase difference means that for any given pair of pulsars, for which the local Axion density is equal, 
range in relative amplitude from zero to double the amplitude of each separately, as shown by the  blue shaded area in Figure 2.  
Furthermore, for well separated pulsars (on a scale greater then the de-Broglie scale) the timing amplitude range can be enhanced 
by another factor of two because of the de-Broglie scale interference that fully modulates the local density about the mean level, shown in Figure 2.

\begin{figure}[t]
\centering
\includegraphics[width=\columnwidth]{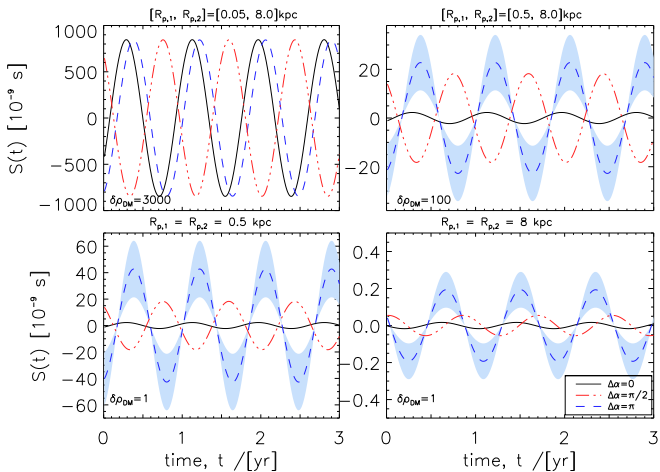}
\caption{Predicted timing signal between pulsar pairs, for a light scalar field of 
$m_B \sim 0.8\times10^{-22}$ eV.  In the top panels  the relative timing signal
between local pulsars at $8$ kpc, and pulsars close to the  Galactic center at a radius of 50 and 500 pc.
In the bottom left and right panels both members of the pair are located  at the 
same Galactocentric radius of $0.5$ and $8$ kpc, 
with relative phases chosen indicated in the inset box, bottom right,
illustrating the signal can cancel in such cases. 
The shaded regions indicate how the density modulation on the de-Broglie scale can enhance or
diminish the pairwise relative timing signal.}\label{fig2}
\end{figure}

Ideally, the relative pulsar timing may not have to rely on being referred to any Earth clock for 
simultaneous observations of different pulsars through the same telescope, or for telescopes that are 
highly synchronous, with the advantage that any vagaries in precision of Earth clocks cancel. So here we calculate the relative timing amplitudes for pairs of pulsars $S(t)$:
\begin{eqnarray}\label{signal}
S(t) &= &\Delta t_1(t) - \Delta t_2(t)=\frac{1}{\omega}\biggl[\Psi({\bf x_1})\sin\left(\omega t + \alpha'_1\right) -\nonumber\\ 
&-& \Psi({\bf x_2})\sin\left(\omega t +\alpha'_2\right)\biggr], 
\end{eqnarray}
where we have defined $\alpha'_i=2\alpha_i- \frac{\omega D_i}{c}$.
To illustrate this difference we calculate the relative timing signal in Figure \ref{fig2} between local pulsars, and
pulsars close to the  Galactic center at a radius of 50 and 500 pc,  fixing the boson mass to $m_B \sim 0.8\times10^{-22}$ eV.
 In this case,  the relative timing amplitude is dominated by the pulsar closer to the Galactic center and it can reach 
an amplitude  of the order of 600 ns. While taking the pulsars at approximately the same distance from the Galactic center,
the relative timing amplitude signal strongly depends on the phase of the pulse, 
canceling when the signals as seen from Earth are in phase.

For convenience, we define ratio of the DM density at the locations of the two pulsars as
\begin{equation}
\delta\rho_{DM}=\left(\frac{\Psi({\bf x_1})}{\Psi({\bf x_2})}\right)=\frac{\rho_{DM}({\bf x_1})}{\rho_{DM}({\bf x_2})}.
\end{equation}

To estimate the detectability of forthcoming pulsar timing arrays to the Axion oscillation we compute the average square signal over all the phases 
\begin{equation}
\sqrt{\langle S^2(t)\rangle}= \frac{\sqrt{2}}{2\omega}\sqrt{\Psi({\bf x_1})^2+\Psi({\bf x_2})^2}, 
\end{equation}
and relate it to the GW strain, and since the $\Psi({\bf x})$ amplitude of the oscillation only depends on the Axion density we can always re-write $\Psi({\bf x_2})$ in term of $\Psi({\bf x_1})$
\begin{equation}\label{eq:hc}
h_c= \frac{\sqrt{6}}{2}\Psi({\bf x_2})\sqrt{1+\delta\rho_{DM}^2}.
\end{equation}

\begin{figure}[t]
\centering
\includegraphics[width=8.6cm]{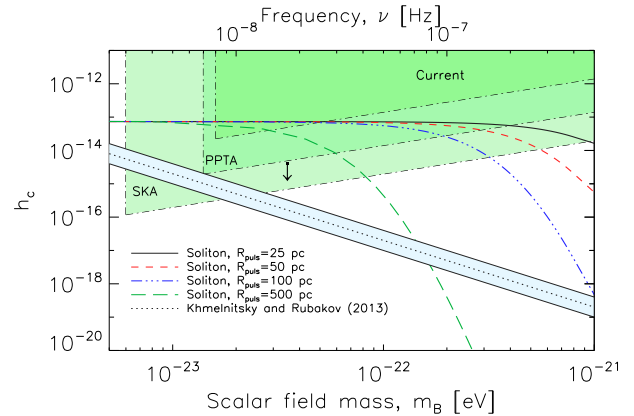}
\caption{Characteristic strain measured between pairs of pulsars, with Galactocentric radii chosen as in Figure 2 and compared with
the expected sensitivities from the current and forthcoming PTA experiments (adapted from \cite{Sesana2010, Khmeinitsky:2013lxt}), 
with the corresponding  oscillation frequency shown above. We highlight the relatively high signal strength 
we calculate for pulsars within the Galactic soliton region as a function of Axion mass in a series of curves, 
demonstrating that this signal is already detectable for central Galactic pulsars. For comparison, the diagonal 
dotted line is the prediction obtained by \cite{Khmeinitsky:2013lxt} for local pulsars, 
assuming a smooth density distribution, with the blue shaded region representing the wider range we predict that includes our 
de-Broglie scale DM density modulation. The local upper limit obtained by \cite{Porayko2014} is also shown (black 
square with arrow).}
\label{fig3}
\end{figure}

 We compute the characteristic amplitude shown in Figure \ref{fig3}, 
highlighting the relatively strong signal expected for pulsars 
within $\sim0.5$ kpc of the Galactic center, 
We also over-plot the confidence regions for the current 
results from Pulsar Timing Array (PTA), Parkes Pulsar Timing Array (PPTA), 
Square Kilometre Array (SKA)  experiments to allow a direct comparison
with expected sensitivities of these new surveys \cite{Sesana2010},  
where even one such central pulsar can provide a sufficient timing 
residual to test for the presence of light axionic dark matter. 


\section{Discussion and Conclusions}

We have considered Compton and de-Broglie scale modulations within the Axionic interpretation of dark 
matter and examined their combined effect on Pulsar timing. For our Galaxy we estimate the de-Broglie scale 
is approximately 100 times larger than the Compton scale, corresponding to $\simeq 150pc$ for the Milky Way, with 
the favoured axion mass of 
$m_B\simeq 10^{-22}eV$. Within virialized halos our simulations reveal that the density distribution is fully 
modulated on the de-Broglie scale, as shown in Figure 1, with a dense soliton at the center of radius $\simeq 150pc$ 
where the pulsar timing effect is strong. Compton oscillation will be coherent within de-Broglie  sized patches, becoming unrelated on larger scales. We have made self consistent predictions for pulsar timing residuals including this spatial dependence revealed in our $\psi_{DM}$ simulations \cite{Schive2014}. 

The de-Broglie interference is most conspicuous by the formation of central soliton on the de-Broglie scale, 
representing a stable, time-independent ground state, where the pulsar timing residuals are expected to be 2-3 orders of magnitude higher than those imprinted on local pulsars, due to the relatively high central 
density of the soliton, that much exceeds in density a corresponding NFW profile.  Such central millisecond 
pulsars are expected in large numbers within the bulge and near the galactic center 
\cite{Figer2004, Calore2016}, and can account for the GeV gamma-ray excess \cite{Abazaijan2011, Abazaijan2014} 
and are being searched with some success \cite{Johnston2006, Deneva2009}. 
Detection will be compromised within the inner 100 pc where the high plasma density is high and the ISM contains small scale irregularities, causing dispersion of pulse arrival time and pulse smearing, although ``corridors" of lower scattering 
may be evident  \cite{Macquart2015, Schnitzeler2016}.
The smearing is a low-pass filter making millisecond pulsars undetectable at low frequency, but decreases rapidly at
high frequency ($<$10Ghz) as  $\nu^{-4}$. Blind searches are not yet very practical at these frequencies with single
dishes and the pulse amplitude is typically relatively weak, but soon the SKA will be efficiently capable of detecting
the putative population of central millisecond pulsars and comprehensive searches are already underway 
\cite{Macquart2015, Bhakta2017}. 

 In terms of the central potential of the Galaxy, the stellar bar dominates, but the signature of this soliton 
 feature is expected dynamically on a scale of $\lesssim 150/m_B$ pc, which lies between the scale length of the
 stellar bulge (1 kpc) and the smaller pc scale region of influence of the central Milky Way black hole. Careful dynamical modeling by Portail et al. \cite{Portail2017} has recently uncovered a central shortfall of $2\times10^9 \msun$ 
 of ''missing matter" which may help account for the 100 km/s motion of stars within the central $\simeq 120$ pc of
 the galaxy \cite{Kruijssen2015, Henshaw2016, Schonrich2015}
 and which we aim to examine in the context of the near spherical soliton potential that we predict here.

 By using pulsars spread over a wide range of Galactocentric radius we may detect the radial dependence of 
pulsar timing ampliude on the dark matter density, beyond current limits claimed 
in \cite{Porayko2014} and pulsar binary resonance \cite{Blas2017}. This dependence helps distinguish the 
monotone Compton scale modulation we seek from an isotropic, stochastic GW background that has broad band variance.   In making such pairwise measurements we have 
stressed that only relative timing variations can be detected as all clocks are modulated 
on the same Compton frequency scale including Earth clocks. Furthermore, we can anticipate a relatively large 
variance of such timing residuals  because of the large density modulation from de-Broglie scale 
interference. The timing residual will absent for Pulsars at density minima but can be 
enhanced by 100\% for pairs that lie near density maxima and located out of phase with respect 
to the Compton oscillation. This means that we cannot rely on only one pair 
of pulsars when assessing the presence of this Compton frequency, but must examine an ensemble to 
average over the combined Compton and de-Broglie modulation for a statistical detection.

For pulsars that may be detected in local dwarf spheroidal 
galaxies where the entire visible content lies within a solitonic 
core \cite{Schive2014},  the timescale is set by $m_B$ but the timing amplitude 
will by lower, set by the soliton density which scales as 
$M_{halo}^{4/3}$ \cite{Schive2014b}, about a factor of $10^{-3}$ smaller than for the 
Galactic center, close to the current PTA capability limit.

\vspace*{1cm}
IDM acknowledges support from University of the Basque Country program
``Convocatoria de contrataci\'{o}n para la especializaci\'{o}n de personal 
investigador doctor en la UPV/EHU 2015", and from the Spanish Ministerio de
Econom\'{\i}a y Competitividad through research project FIS2010-15492,
and the Basque Government research project IT-956-16.
RL is supported by the Spanish Ministry of Economy and Competitiveness through research projects
FIS2010-15492 and Consolider EPI CSD2010-00064, and the
University of the Basque Country program UFI 11/55. IDM and RL also 
acknowledges the support from the COST Action CA1511 Cosmology and Astrophysics 
Network for Theoretical Advances and Training Actions (CANTATA).
TJB acknowledges generous hospitality from the Institute for Advanced Studies in Hong Kong and helpful 
conversations with Nick Kaiser and Kfir Blum. SHHT is supported by the CRF Grant HKUST4/CRF/13G and 
the GRF 16305414 issued by the Research Grants Council (RGC) of the Government of the Hong Kong SAR.
The GPU cluster donated by Chipbond Technology Corporation, with which this work is conducted, 
is acknowledged. This work is supported in part by the National Science Council of Taiwan under 
the grants NSC100-2112-M-002-018-MY3 and NSC99-2112-M-002-009-MY3.

\end{document}